\documentclass{article}
\usepackage{spconf,amsmath,epsfig}
\usepackage{amssymb}
\let\OLDthebibliography\thebibliography
\renewcommand\thebibliography[1]{
  \OLDthebibliography{#1}
  \setlength{\parskip}{0pt}
  \setlength{\itemsep}{0pt plus 0.3ex}
}

\begin{document}\sloppy

\def\x{{\mathbf x}}
\def\L{{\cal L}}

\title{3D-VFD: A Victim-free Detector against 3D Adversarial Point Clouds}
%
\name{Jiahao Zhu\textsuperscript{1}, Huajun, Zhou\textsuperscript{1}, Zixuan Chen\textsuperscript{1}, Yi Zhou\textsuperscript{2}, Xiaohua Xie\textsuperscript{1}}
\address{\textsuperscript{1}School of Computer Science and Engineering, Sun Yat-sen University, China\\
\textsuperscript{2}Zhongshan School of Medicine, Sun Yat-sen University, China\\
\{zhujh59,zhouhj26,chenzx3\}@mail2.sysu.edu.cn,\{zhouyi,xiexiaoh6\}@sysu.edu.cn}

\maketitle

\begin{abstract}
3D deep models consuming point clouds have achieved sound application effects in computer vision. However, recent studies have shown they are vulnerable to 3D adversarial point clouds. In this paper, we regard these malicious point clouds as 3D steganography examples and present a new perspective, 3D steganalysis, to counter such examples. Specifically, we propose 3D-VFD, a victim-free detector against 3D adversarial point clouds. Its core idea is to capture the discrepancies between residual geometric feature distributions of benign point clouds and adversarial point clouds and map these point clouds to a lower dimensional space where we can efficiently distinguish them. Unlike existing detection techniques against 3D adversarial point clouds, 3D-VFD does not rely on the victim 3D deep model's outputs for discrimination. Extensive experiments demonstrate that 3D-VFD achieves state-of-the-art detection and can effectively detect 3D adversarial attacks based on point adding and point perturbation while keeping fast detection speed.
\end{abstract}
\section{Introduction}
\label{sec:intro}
3D deep models directly consuming point clouds, e.g. PointNet \cite{Qi_2017_CVPR} and PointNet++ \cite{pointnet++}, have been extensively used in 3D shape classification, 3D object locating and tracing, and 3D object segmentation \cite{3Dpointsurvey}. However, studies in recent years have demonstrated that such models lack robustness against 3D adversarial point clouds \cite{SIA, ITA}, which may cause untold damage to safety-critical applications, such as autopilot and autonomous driving. 

Current 3D adversarial point clouds can be generated by point shifting, point adding, or point removing. Fig. \ref{fig1} enumerates several visual results generated by the three types of 3D adversarial attacks. We can see that adversarial examples in 3D settings are no longer limited to the concept of \emph{imperceptible perturbation} \cite{SIA,ITA,Xiang_2019_CVPR,yang2021adversarial} but are placed in a more open environment. Attackers in such an environment will fully use point clouds' irregularity and unordered data structure and then adopt more radical methods, such as deleting \cite{yang2021adversarial,Saliency, Wicker_2019_CVPR} or adding points \cite{Xiang_2019_CVPR,shapeattack}, to create adversarial point clouds. These methods are often easier to deploy in the physical world and thus more aggressive with 3D deep models \cite{tu2020physically}.

As the damage caused by adversarial attacks on 3D deep models is substantial, defensive countermeasures arise at the historic moment. Initially, scholars followed a ``detect-and-reject'' strategy to counter 3D adversarial attacks \cite{yang2021adversarial}. Such defense does not involve modification of models and inputs, so we consider it passive defense. Later, input preprocessing \cite{DUP, Liushape}, network architecture modification \cite{IF, GVG}, and adversarial training \cite{Liushape} have been spreadly used to mitigate adversarial attacks. In recent years, there has been a rise in research on building certifiably robust 3D deep models to resist adversarially shifted, added, or removed points \cite{PointGuard}. Unlike passive defense, these defense methods will alter input data or deep models, so we call them active defense.
\begin{figure}[t]
	\centering	\includegraphics[width=0.5\textwidth]{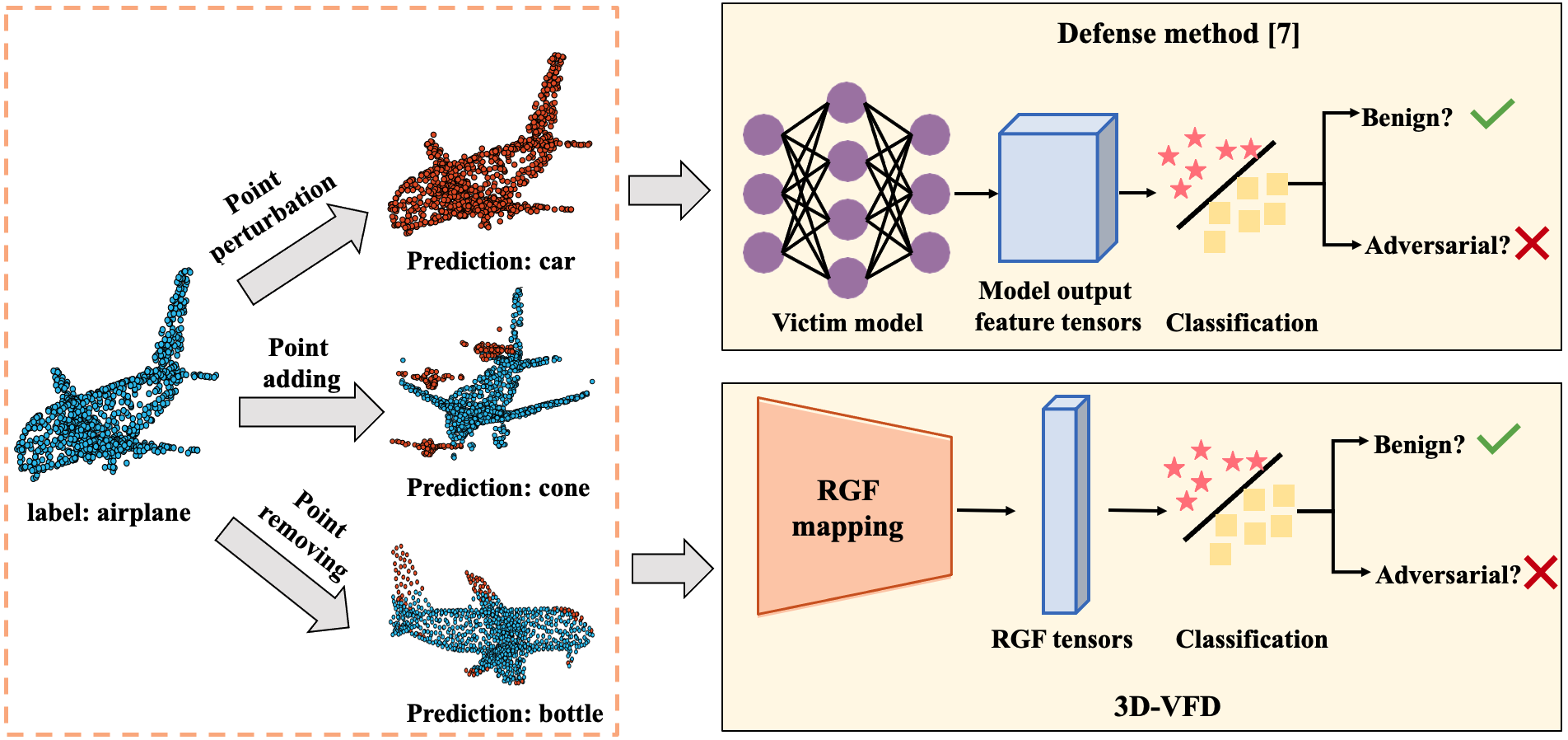}
	\label{fig1}
	\caption{Left half: visualizations of 3D adversarial point clouds based on point perturbation, point adding, and point removing, respectively, where the red points denote the adversarial points. Right half: 3D-VFD and defense method \cite{yang2021adversarial} both follow the ``detect-and-reject'' strategy, but ours does not rely on the victim model's outputs; instead, it attempts to capture the discrepancies between RGF distributions of benign point clouds and adversarial ones.}
 \label{fig1}
 \vspace{-1em}
\end{figure}

While the methods above achieve fairly impressive defense results, they still have potential shortcomings:  1) input processing-based defenses will modify both malicious point clouds and benign point clouds, which is unfriendly to precision-critical applications; 2) In addition, network architecture modification and adversarial training involve network retraining, which is costly to enterprises, especially those large ones; 3) For the existing detection based method \cite{yang2021adversarial}, its detection relies heavily on the victim model's outputs, which is reflected in Fig. \ref{fig1}.

In fact, \cite{goodfellow2015explaining} has implied that adversarial attacks can be regarded as a sort of ``accidental steganography''. Inspired by this, in this paper, we take a new perspective, 3D steganalysis, on 3D adversarial point cloud defense and propose 3D-VFD, a victim-free 3D adversarial point cloud detector. Unlike the active defenses mentioned above, 3D-VFD neither preprocesses input point clouds nor modifies victim models' parameters and architectures. Contrary to the passive one \cite{yang2021adversarial}, 3D-VFD is able to distinguish the malicious point clouds from benign ones without the knowledge of victim models, as shown in Fig \ref{fig1}. 3D-VFD consists of a residual geometric feature (RGF) mapping module and a classification module. The first module applies a feature extractor with a siamese structure to get the RGF tensors of benign and adversarial point clouds, mapping them to a lower dimensional space where we can distinguish them efficiently. The second module takes the obtained feature tensors and judges whether they are adversarial examples. The main contributions of this work are threefold:
\begin{itemize}
\item We present a new defense perspective, 3D steganalysis, for 3D adversarial attacks and experimentally prove it can be applied to detect 3D adversarial point clouds.
    \item We propose 3D-VFD, which consists of the RGF mapping module and classification module. Unlike previous defenses, it requires less prior knowledge and fewer channel resources.
    \item Experiments demonstrate that 3D-VFD achieves state-of-the-art detection performance and effectively detects adversarial point clouds produced by point perturbation and point adding while keeping fast detection speed.
\end{itemize}
\section{Proposed Method}
\label{sec:pagestyle}
\begin{figure}[t]
	\centering 
\includegraphics[width=\linewidth]{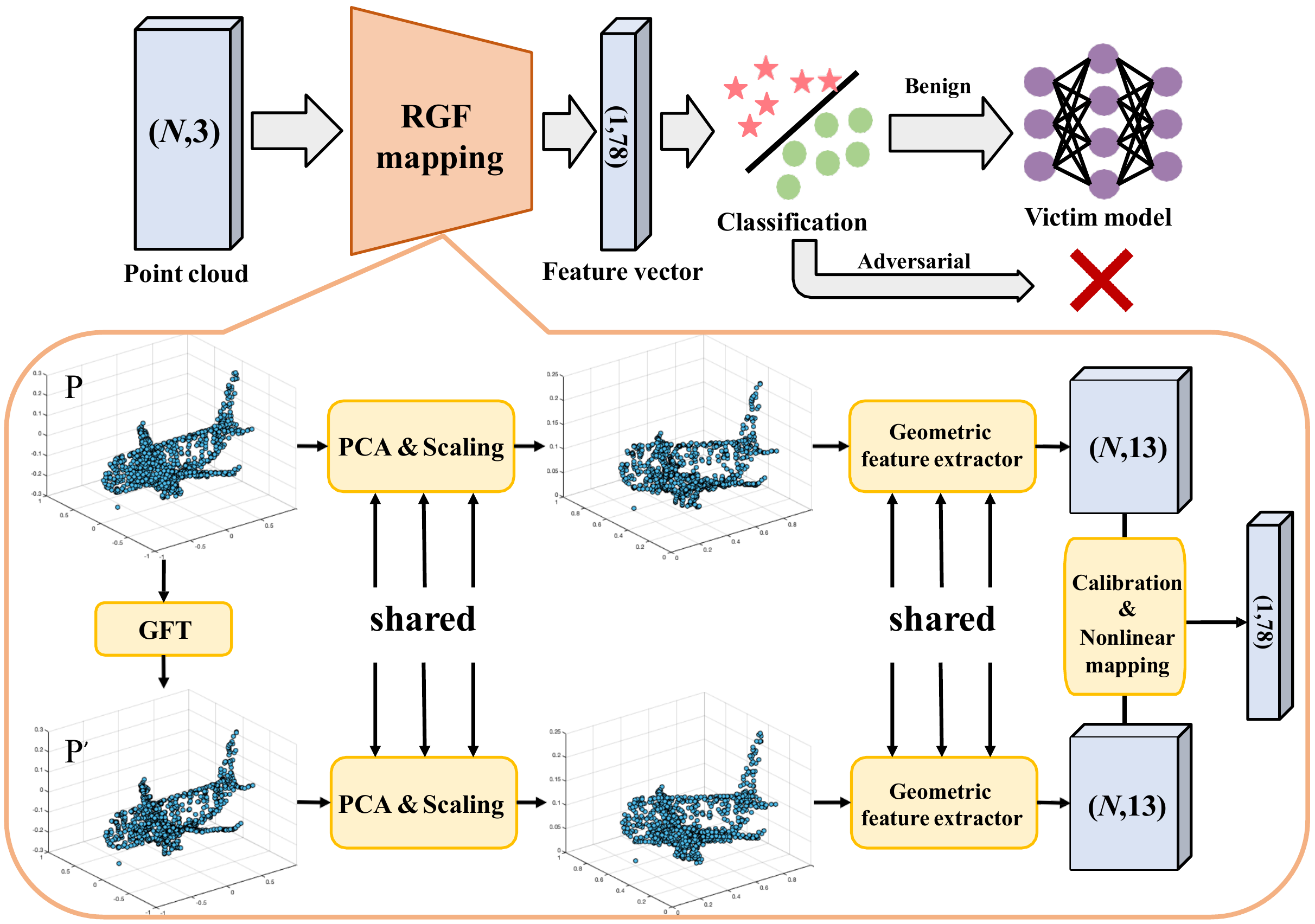}
	\caption{Workflow of 3D-VFD (upper half). The bottom half portrays details of the RGF mapping with a siamese structure.}
 \label{fig2}
\end{figure}

3D-VFD consists of two basic modules, RGF mapping and classification, as shown in Fig. \ref{fig2}. The former is defined by $\mathcal{F}: \mathbb{R}^{N\times3}\rightarrow \mathbb{R}^{78}$, which takes a point cloud $\mathbf{P}=[\boldsymbol{p}_1,\cdots,\boldsymbol{p}_N]$ as input and outputs a 78-dimensional feature vector. The specific implementation of the RGF mapping with a siamese structure is also given in Fig. \ref{fig2}. The obtained feature vectors will be trained with the off-the-shelf classification learners in the second module. In practice, our defense can be deployed by a collection of the designed detectors trained for various mainstream 3D adversarial attacks.
\subsection{Implementation of RGF Mapping}
\subsubsection{GFT, PCA, and Scaling}
In this phase, we first generate a reference counterpart $\mathbf{P}'$ of $\mathbf{P}$ for the subsequent feature calibration step with the Graph Fourier Transform (GFT) based smoothing \cite{GFT}. Since point clouds do not contain connectivity information, we cannot directly adopt GFT to smooth point clouds. To tackle this problem, we first assume that each point of $\mathbf{P}$ is connected with its $K_g$ nearest neighbors $N(\boldsymbol{p}_i,K_{g})$, and the adjacency matrix can be defined by
\begin{equation}\label{eq1}
	\mathbf{A}_{ij} = \begin{cases}
		e^{-\frac{(\boldsymbol{p}_i-\boldsymbol{p}_j)^{\top}(\boldsymbol{p}_i-\boldsymbol{p}_j)}{2\alpha}}, & \boldsymbol{p}_j \in N(\boldsymbol{p}_i,K_{g}) \\
		0,& otherwise \\
	\end{cases},
\end{equation} 
$\mathbf{A}$ is probably not a symmetric matrix, which may result in the Laplacian matrix not being able to be eigendecomposed. According to Eq. (\ref{eq1}), when $\mathbf{A}_{ij} \ne 0$, $\mathbf{A}_{ji}$ is either equal to $\mathbf{A}_{ij}$ or not. Therefore, we can transform $\mathbf{A}$ efficiently as below to make itself symmetrical. 
\begin{equation}\label{tr1}
	\mathbf{A} = \mathbf{A} + \frac{1}{2}abs(\mathbf{A}-\mathbf{A}^\top-abs(\mathbf{A}-\mathbf{A}^\top))
\end{equation} 
where $abs()$ is an elementwise absolute value operator. We do not strictly require that each point can only have $K_g$ neighbors; instead, the symmetrized $\mathbf{A}$ indicates that each point of $\mathbf{P}$ is connected with \emph{at least $K_g$ nearest neighbors}, denoted by $\widehat{N}(\boldsymbol{p}_i,K_g)$. The normalized Laplacian matrix $\mathbf{L}=\mathbf{D}^{1/2}(\mathbf{D}-\mathbf{A})\mathbf{D}^{1/2}$, where $\mathbf{D}$ is the diagonal matrix following from $\mathbf{A}$.
Note that setting $\alpha$ too large/small will cause $\mathbf{A}_{ij}$ to approach 1/0. To avoid such issues while guaranteeing the symmetry of $\mathbf{L}$, we set 
\begin{equation}\label{alpha}
	\alpha=\frac{1}{N|\widehat{N}(\boldsymbol{p}_i,K_g)|}\sum_{i=1}^{N}\sum_{\boldsymbol{p}_j \in \widehat{N}(\boldsymbol{p}_i,K_{g})}(\boldsymbol{p}_i-\boldsymbol{p}_j)^{\top}(\boldsymbol{p}_i-\boldsymbol{p}_j).
\end{equation}
Next we need to calculate the eigenvector matrix $\mathbf{Q}=[\boldsymbol{q}_1,\cdots,\boldsymbol{q}_N]$ of $\mathbf{L}$, where $\boldsymbol{q}_i \in \mathbb{R}^{N\times1}$ and they are arranged according to eigenvalue ascent.
The sharp features usually lie in the low dimensional subspace of $\mathbf{Q}$, while noise is the opposite. Based on this, we can obtain the smoothed counterpart $\mathbf{P}'$ of $\mathbf{P}$ by
\begin{equation}
	\mathbf{P}'=\mathbf{P}[\boldsymbol{q}_1,\cdots,\boldsymbol{q}_t], t<N.
\end{equation}
How to determine the values of $t$ and $K_{g}$ will be explored in Section 4. Finally, we transform $\mathbf{P}$ and $\mathbf{P}'$ via Principal Component Analysis (PCA) and then normalize them into unit cubes separately, whose respective center points are $[0.5, 0.5, 0.5]^{\top}$. 

\subsubsection{Geometric Feature extraction}
\emph{We do not consider deep learning in our geometric feature extraction due to the difficulty in distinguishing 3D objects with similar appearance using deep networks \cite{ZhouS}}. Nonetheless, the feature set based on the following hand-crafted geometric features shows acceptable distinguishing ability.

\textbf{Normal based features.} We adopt the traditional method \cite{normale} to obtain the point normals of $\mathbf{P}$. Firstly, we compute the covariance matrix w.r.t. $\boldsymbol{p}_i$ as 
\begin{equation}
	\mathbf{C}_i = \sum_{\boldsymbol{p}_i \in N(\boldsymbol{p}_i,K_{v})}(\boldsymbol{p}_j-\boldsymbol{p}_i)(\boldsymbol{p}_j-\boldsymbol{p}_i)^{\top}
\end{equation}
The exterior product of the two eigenvectors corresponding to two maximum eigenvalues of $\mathbf{C}_i$ is regarded as the point normal of $\boldsymbol{p}_i$. Finally, we can get two normal tensors $\mathbf{N}$ and $\mathbf{N}'$ w.r.t. $\mathbf{P}$ and $\mathbf{P}'$ respectively, where $\mathbf{N}=[{\boldsymbol{n}_1,\cdots,\boldsymbol{n}_N}] \in \mathbb{R}^{3\times N}$. $K_{v}$ will be discussed in Section 4.

\textbf{Curvature based features.} We take a simple approach to estimating the curvature at $\boldsymbol{p}_i$. First of all, we construct a projection matrix w.r.t. $\boldsymbol{p}_i$ by
\begin{equation}
	\boldsymbol{\rm{Pr}}_i = \boldsymbol{\rm{I}}-\boldsymbol{n}_i\cdot\boldsymbol{n}^\top_i,
\end{equation}
where $\mathbf{I}$ is an identity matrix. Then we project the point normals of $\boldsymbol{p}_i$'s $K_{c}$ nearest neighbors into the tangent plane they determine by $\boldsymbol{n}^{proj}_j=\boldsymbol{\rm{Pr}}_i \boldsymbol{n}_j$. The set of $K_{c}$ will be discussed in Section 4. Next, we calculate the covariance matrix of these $\boldsymbol{n}^{proj}_j$ and regard its maximum eigenvalue $C_1$ and the minimum eigenvalue $C_2$ as the principal curvature at $\boldsymbol{p}_i$. We consider the Gaussian curvature, mean curvature, and curvature ratio as features, which are respectively defined by
\begin{equation}
	\begin{aligned}
		&GC_i = C_1\times C_2, \ MC_i = \frac{C_1+C_2}{2} \\
		&CR_i = \frac{\mathop{\rm{min}}(abs(C_1),abs(C_2))}{\mathop{\rm{max}}(abs(C_1),abs(C_2))}
	\end{aligned}
\end{equation}
Finally, we can get three curvature based tensors $\mathbf{GC}, \mathbf{MC}$, $\mathbf{CR} \in \mathbb{R}^{N\times1}$. Similarly, we can generate $\mathbf{GC}', \mathbf{MC}'$, and $\mathbf{CR}'$ for $\mathbf{P}'$.

\textbf{Normal voting tensor (NVT) based features.} NVT based features show exceptional performance in 3D mesh steganalysis \cite{NVT2021} because they can well capture local correlation features. Inspired by this, we consider NVT based features for 3D adversarial example detection. \cite{NVT2021} proposed a NVT model applicable for point clouds, but it does not work in our defense method due to its crude point normal estimation. To make more effective use of NVT, we propose the following tensor model.
\begin{equation}\label{eq8}
	\mathbf{T}_i = \sum_{\boldsymbol{n}_j\in N(\boldsymbol{n}_i,K_{n})}\mu_{i,j}\boldsymbol{n}_j\cdot \boldsymbol{n}_j^{\top}, \ i=1,\cdots,N,
\end{equation}
where $\mu_{i,j} = e^{-\frac{(\boldsymbol{n}_i-\boldsymbol{n}_j)^{\top}(\boldsymbol{n}_i-\boldsymbol{n}_j)}{2\beta}}$ and it is also sensitized to point sampling. To avoid $\mu_{i,j}$ approaching 0/1, we set
\begin{equation}
	\beta = \frac{1}{K_{n}}\sum_{\boldsymbol{n}_j\in N(\boldsymbol{n}_i, K_{n})}(\boldsymbol{n}_i-\boldsymbol{n}_j)(\boldsymbol{n}_i-\boldsymbol{n}_j)^\top.
\end{equation}
Since $\mathbf{T}_i$ is positive semidefinite, without losing generality, let $\lambda_{i1} \ge \lambda_{i2} \ge \lambda_{i3}\ge0$ be the three eigenvalues of $\mathbf{T}_i$. According to \cite{VTtheory}, $\mathbf{T}_i$ can be decomposed into three tensors that describe a stick, a plate, and a ball respectively. Coefficients corresponding to the three tensors, i.e., $\lambda_{i1}-\lambda_{i2}$, $\lambda_{i2}-\lambda_{i3}$, and $\lambda_{i3}$, will be used as features. Finally, we can get three feature tensors $\boldsymbol{\lambda}_1=[\lambda_{11}-\lambda_{12},\cdots,\lambda_{N1}-\lambda_{N2}]^\top$, $\boldsymbol{\lambda}_2=[\lambda_{12}-\lambda_{13},\cdots,\lambda_{N2}-\lambda_{N3}]^\top$, and $\boldsymbol{\lambda }_3=[\lambda_{13},\cdots,\lambda_{N3}]^\top$ for $\mathbf{P}$. Accordingly, we can get $\boldsymbol{\lambda}'_1$, $\boldsymbol{\lambda}'_2$, and $\boldsymbol{\lambda}'_3$ for $\mathbf{P}'$. Additionally, we also consider another two neighborhood types for Eq. (\ref{eq8}), namely $N(\boldsymbol{n}_i, 2K_{n})$ and $N(\boldsymbol{n}_i, 3K_{n})$, their respective feature extractions are same as $N(\boldsymbol{n}_i, K_{n})$'s. How to set $K_{n}$ will also be discussed in Section 4.
\subsubsection{Calibration}
Calibration helps make our method focus more on the residual features rather than the point cloud itself, which is one of the reasons why our defense method can successfully discriminate between benign and malicious point clouds. For normal based features, their calibrated features are given by
\begin{equation}
	\mathbf{N}_c = [\mathop{\rm{arccos}}\frac{\boldsymbol{n}_1\cdot \boldsymbol{n}'_1}{||\boldsymbol{n}_1||_2\cdot||\boldsymbol{n}'_1||_2},\cdots,\mathop{\rm{arccos}}\frac{\boldsymbol{n}_N\cdot \boldsymbol{n}'_N}{||\boldsymbol{n}_N||_2\cdot||\boldsymbol{n}'_N||_2}]^\top.
\end{equation}

For curvature based features, their calibrated features are $\mathbf{GC}_c=abs(\mathbf{GC}-\mathbf{GC}')$,  $\mathbf{MC}_c=abs(\mathbf{MC}-\mathbf{MC}')$, and $\mathbf{CR}_c=abs(\mathbf{CR}-\mathbf{CR}')$, respectively. Calibrated counterparts of NVT based features w.r.t. $N(\boldsymbol{n}_i, K_{n})$ are $\boldsymbol{\lambda}^{K_{n}}_c=[abs(\boldsymbol{\lambda}_1-\boldsymbol{\lambda}'_1),abs(\boldsymbol{\lambda}_2-\boldsymbol{\lambda}'_2),abs(\boldsymbol{\lambda}_3-\boldsymbol{\lambda}'_3)]$. Likewise, we can get those w.r.t. $N(\boldsymbol{n}_i, 2K_{n})$ and $N(\boldsymbol{n}_i, 3K_{n})$. The final calibrated feature tensor is given by 
\begin{equation}
	\boldsymbol{\Phi}=[\mathbf{N}_c, \mathbf{GC}_c,\mathbf{MC}_c,\mathbf{CR}_c,\boldsymbol{\lambda}_c^{K_{n}},\boldsymbol{\lambda}_c^{2K_{n}},\boldsymbol{\lambda}_c^{3K_{n}}] \in \mathbb{R}^{N\times13}.
\end{equation}
\subsubsection{Nonlinear Mapping}
To obtain more valuable features, we take the nonlinear mapping in \cite{NVT2021} on $\boldsymbol{\Phi}$, which has been widely used in various 3D steganalyzers and been empirically proven effective for 3D steganalysis. We first logarithmize $\boldsymbol{\Phi}$ and then calculate six statistics: minimum, maximum, mean, variance, skewness, and kurtosis, w.r.t each column of the logarithmized $\boldsymbol{\Phi}$. Finally, we can get a 78-dimensional feature vector. 
\subsection{Classification}
After obtaining the 78-dimensional feature vector of a point cloud, we feed it into a binary classifier for the final result. If judged to be a benign example, the point cloud will be input into the victim model immediately; otherwise, it will be discarded. We consider three types of classification learners, namely linear discriminant (LD), support vector machines (SVMs), and Fisher linear discriminant ensemble (FLDE) \cite{FLD2012}. Relevant experiments are given in Section 4.2.
\section{Experiments}
\subsection{Experimental Settings}
We use the normalized ModelNet40 dataset for experiments. The classical 3D deep model PointNet \cite{Qi_2017_CVPR}, trained on the normalized dataset, is chosen as the victim model. Prior work \cite{2DDetection} demonstrated that untargeted attacks are easier to succeed, so the 3D adversarial attacks we detect in the following experiments all pertain to such attack category.
\subsection{Classification Learner Comparison}
We test the performance of LD, SVMs, and FLDE on the adversarial attacks APP (adversarial point perturbation) \cite{Xiang_2019_CVPR}, AIP (adversarial independent perturbation) \cite{Xiang_2019_CVPR}, and PD-SM (point dropping based on saliency maps) \cite{Saliency}, which are based on point shifting, point adding, and point removing respectively. We prepare 1460 pairs of data for each attack, each containing a benign point cloud and its adversarial counterpart, wherein 10\% are used for the test and the remainder as the training set. FLDE is trained with default settings given in \cite{FLD2012}, and SVMs and LD are all provided by Matlab software. Note that for adversarial point clouds generated by AIP or PD-SM, each one's point number is inconsistent with its benign counterpart's. Directly extracting features from them may give our detector a false hint that benign examples contain 1024 points while malicious ones are not. Therefore we resample each benign example such that it contains the same point number as its corresponding adversarial one. In this experiment, we only consider AIP with 32 points added and PD-SM with 50 points dropped, denoted by AIP(+32) and PD-SM(-50), respectively. Moreover, For APP, AIP(+32), and PD-SM(-50), we just set parameters $t=900$, $k_g=5$, $k_v=5$, $K_c=5$, and $K_n=5$ in this experiment. Table \ref{tab1} and Table \ref{tab2} show their respective classification accuracy and training time, respectively. As we can see, FLDE gains an apparent advantage in classification accuracy. While its training speed is not as fast as that of the LD, its training complexity is acceptable compared to the SVMs. In this work, being more concerned with the classification accuracy and considering the fact that FLDE is extensively used in steganalysis, we decided to use FLDE as the classifier for the following experiments.
\begin{table}[t]
\begin{center}
\caption{Test accuracy of SVMs, LD, and FLDE w.r.t. APP, AIP(+32), and PD-SM(-50).} \label{tab1}
\begin{tabular}{c|ccc}
  \hline
   Accuracy (\%)& APP & AIP(+32)& PD-SM(-50)\\
  \hline
  Linear SVM & 79.7& 75.4 & 50.7\\
  Quadratic SVM & 79.7 & 72.9 & 48.9\\
  Gaussian SVM & 78.3& 69.9 & 51.8\\
  LD & 79.6 & 75.5 & 50.5\\
  Quadratic LD & 77.7 & 69.9 & 47.8\\
  FLDE & \textbf{80.3} & \textbf{77.0} &\textbf{52.7} \\
  \hline
\end{tabular}
\end{center}
\vspace{-2em}
\end{table}
\begin{table}[t]
\begin{center}
\caption{Training time of SVMs, LD, and FLDE w.r.t. APP, AIP(+32), and PD-SM(-50).} \label{tab2}
\begin{tabular}{c|ccc}
  \hline
  Time (s)& APP & AIP(+32)& PD-SM(-50)\\
  \hline
  Linear SVM & 16.73 & 12.50 & 11.88\\
  Quadratic SVM & 28.76 & 26.22 & 43.04\\
  Gaussian SVM & 7.40 & 12.16 & 8.99\\
  LD & 2.16  & 5.55 & 4.02\\
  Quadratic LD & \textbf{1.98} & \textbf{3.27} & \textbf{2.70}\\
  FLDE & 6.67 & 7.08 & 7.12\\
  \hline
\end{tabular}
\vspace{-2em}
\end{center}
\end{table}
\begin{figure*}[t]
	\centering
\begin{minipage}{0.19\textwidth}
  \includegraphics[width=\textwidth]{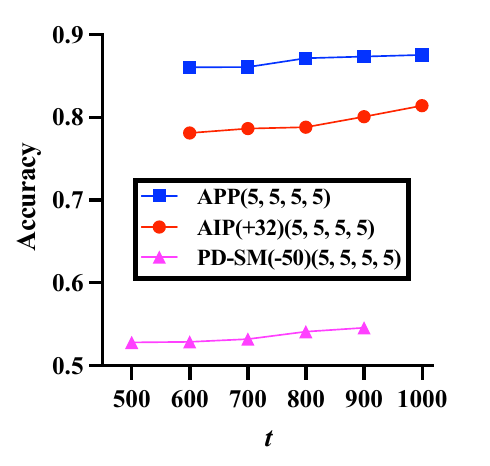}
  \end{minipage}
  \begin{minipage}{0.19\textwidth}
  \includegraphics[width=\textwidth]{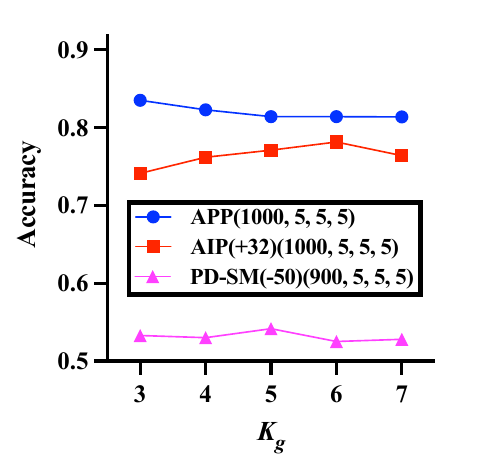}
  \end{minipage}
  \begin{minipage}{0.19\textwidth}
  \includegraphics[width=\textwidth]{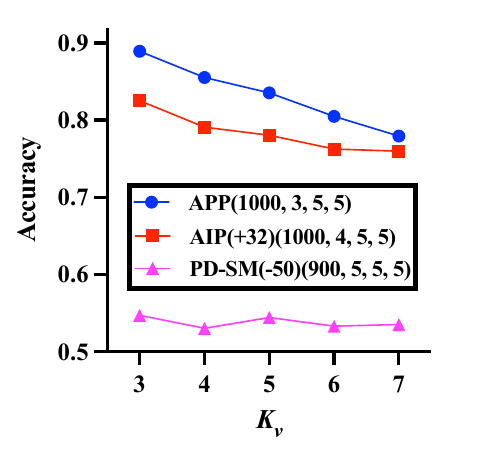}
  \end{minipage}
  \begin{minipage}{0.19\textwidth}
  \includegraphics[width=\textwidth]{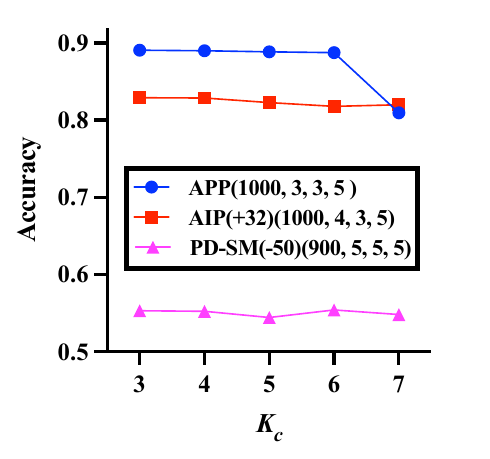}
  \end{minipage}
  \begin{minipage}{0.19\textwidth}
  \includegraphics[width=\textwidth]{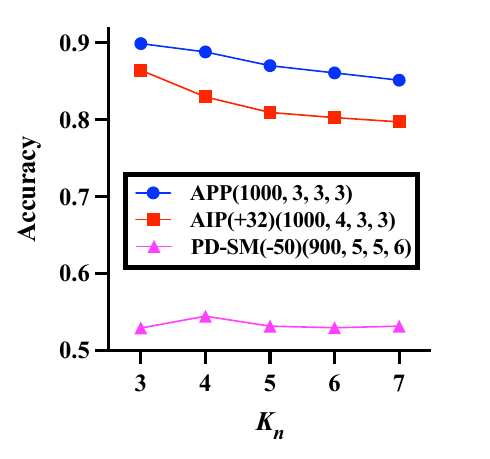}
  \end{minipage}
	\caption{A greedy parameter decision process w.r.t. 3D adversarial point clouds generated by point perturbation method (APP), point adding based method (AIP(+32)), and point removing based method (PD-SM(-50)).}
    \label{fig3}
\end{figure*}
\subsection{Parameter Determination}
As mentioned in Section 3, five parameters may affect our detector's performance, namely $t$, $K_g$, $K_v$, $K_c$, and $K_n$. However, finding their optimal values from the parameter space takes much work. Here we adopt a greedy parameter determination strategy. Before beginning, we need to determine the processing order of the five parameters first. As we know in Section 3, all calibrated features depend on the reference point cloud $P'$ generated by GFT, and thus we give priority to the determination of $t$ and $K_g$. Further, we observe that both curvature based and NVT based features involve calculations related to point normal vectors, so $K_v$ is set in the third order, and $K_c$ and $K_n$ are placed last for processing. Finally, we determine the five parameters in the following order: $t\rightarrow K_g\rightarrow K_v \rightarrow K_c \rightarrow K_n$. 

We will exploit the parameter processing order and 3D-VFD's detection accuracy on APP, AIP(+32) and PD-SM(-50) to roughly estimate the five parameters. Specifically, we first give the parameter initialization (1000, 5, 5, 5, 5) for APP and AIP(+32) and (900, 5, 5, 5, 5) for PD-SM(-50), wherein the neighborhood-related parameters are required to be greater than or equal to 3 in this work. Next, in each step of the parameter determination, we will pick out the best-performing parameter value for the next test. The training set and test set are the same as those given in Section 4.2. As shown in Fig. \ref{fig3}, 3D-VFD favors larger $t$ and smaller neighborhood related parameters when detecting the perturbation based method APP. The detector exhibits the same preference except for $K_g$ when detecting the point adding based method. However, for PD-SM(-50), our detector exhibits no clear parameter preference except for $t$. Note that bigger is not always better for $t$ because, for example, when $t=N$, $\emph{P}$ and $\emph{P'}$ will be almost the same, then the performance of 3D-VFD will drop sharply. Here we empirically give the parameter settings for the subsequent experiments: perturbation based methods ($N$-20, 3, 3, 3, 3), point adding based methods ($N$-20, 6, 3, 3, 3), and point-removing based methods ($N$-20, 5, 5, 5, 4). 
\begin{figure}[t]
	\centering
    \begin{minipage}{0.2\textwidth}
    \includegraphics[width=\textwidth]{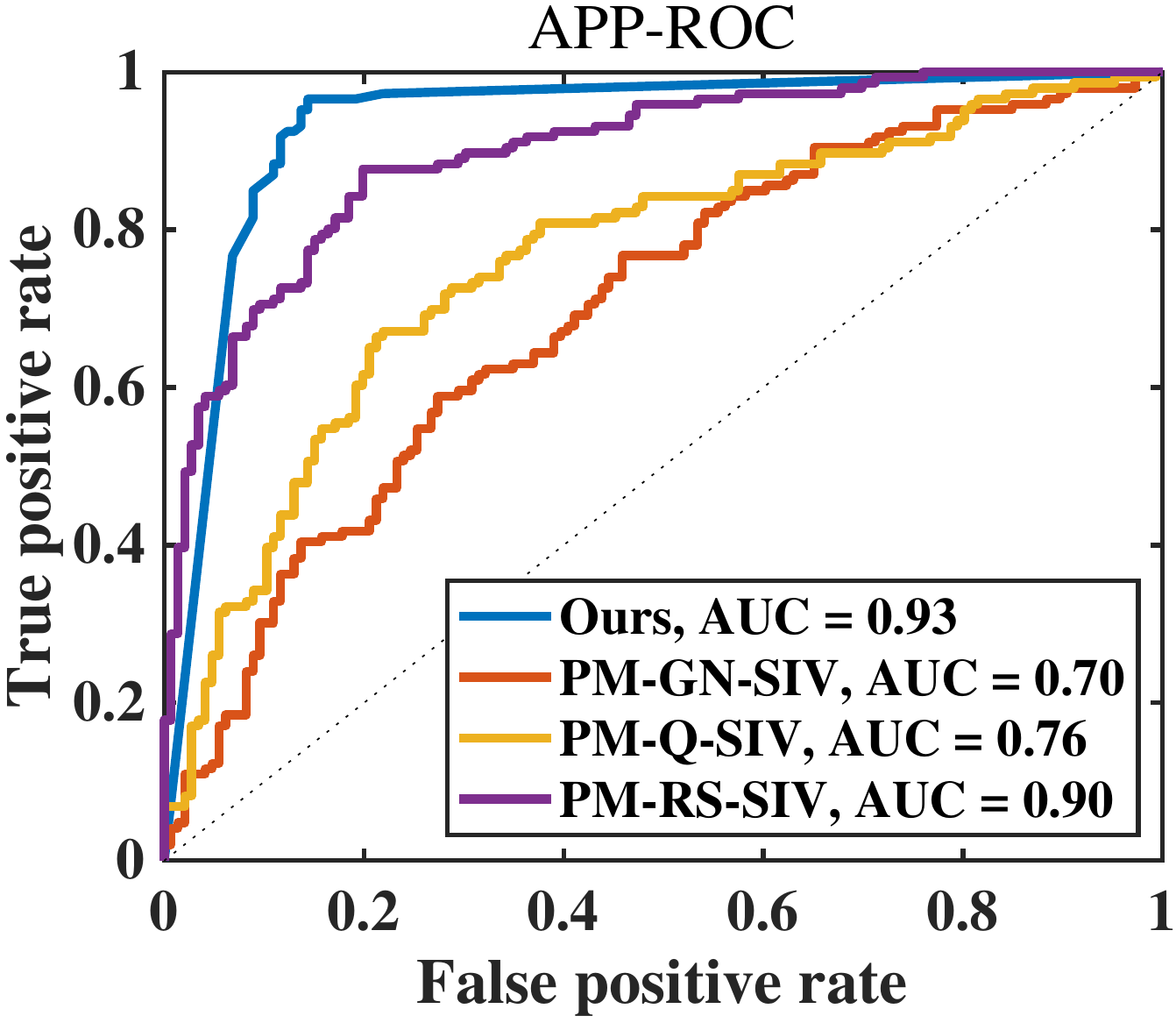}
    \end{minipage}\hspace{+1em}
    \begin{minipage}{0.2\textwidth}
    \includegraphics[width=\textwidth]{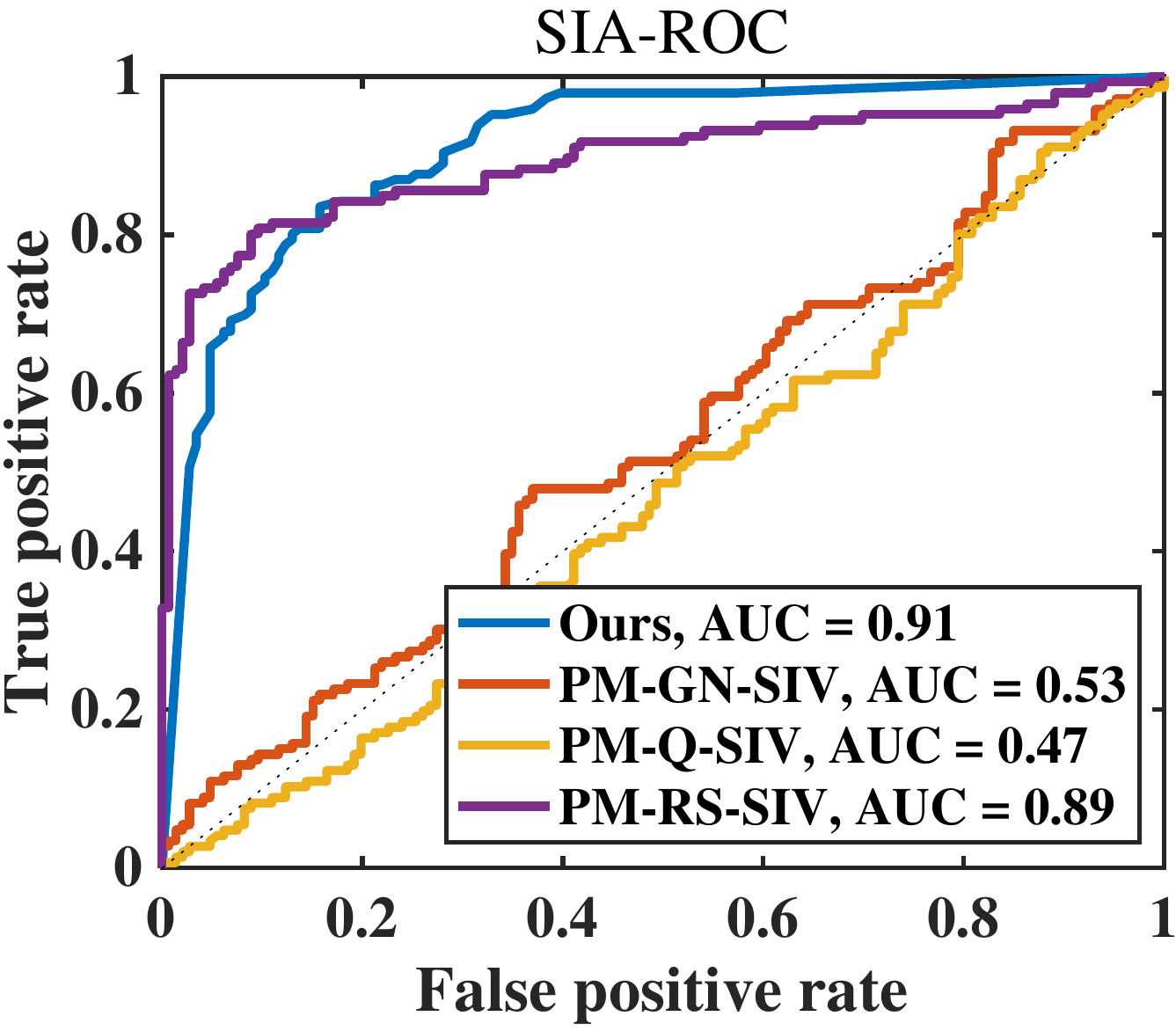}
    \end{minipage}
    \begin{minipage}{0.2\textwidth}
    \includegraphics[width=\textwidth]{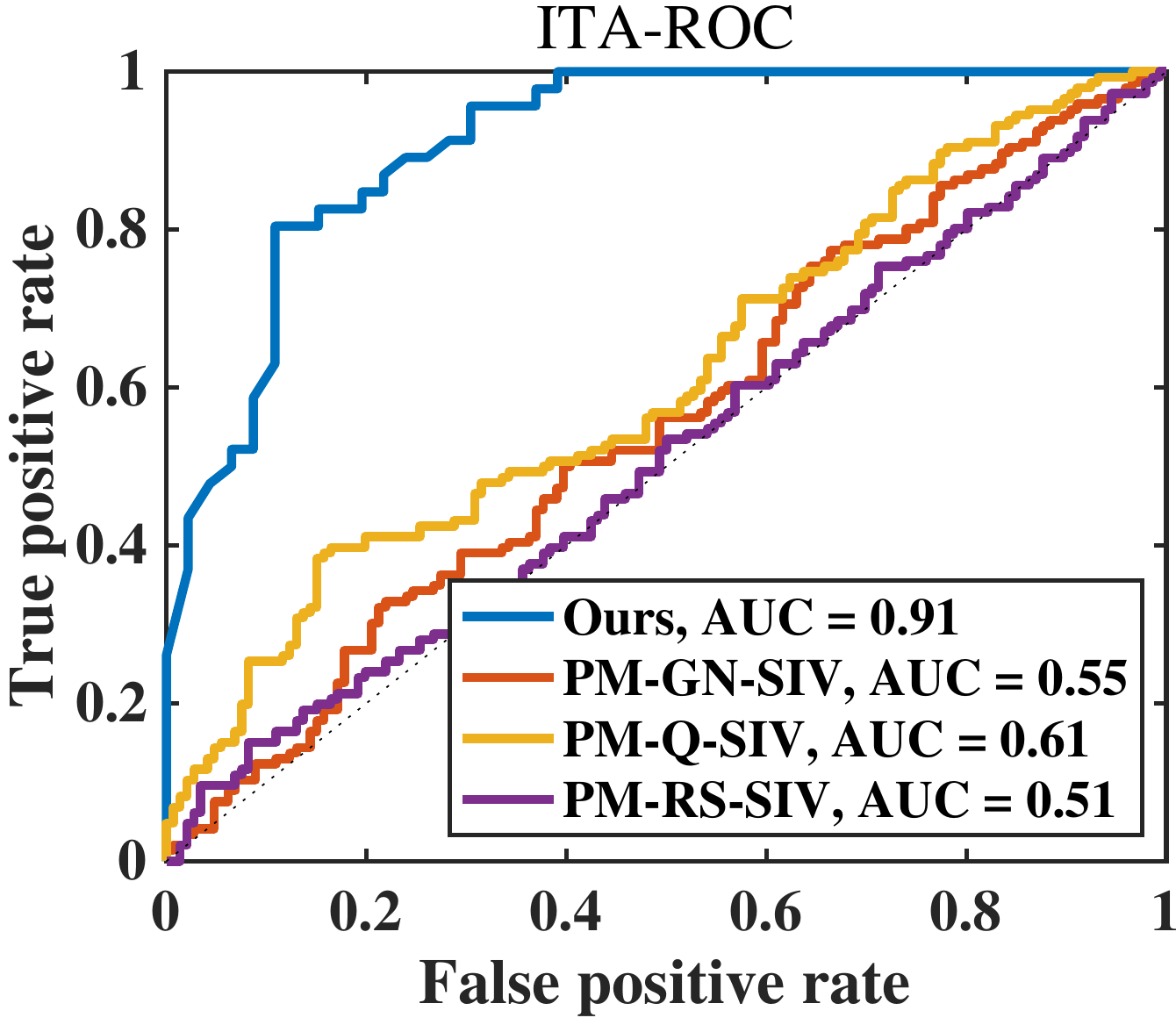}
    \end{minipage}\hspace{+1em}
    \begin{minipage}{0.2\textwidth}
    \includegraphics[width=\textwidth]{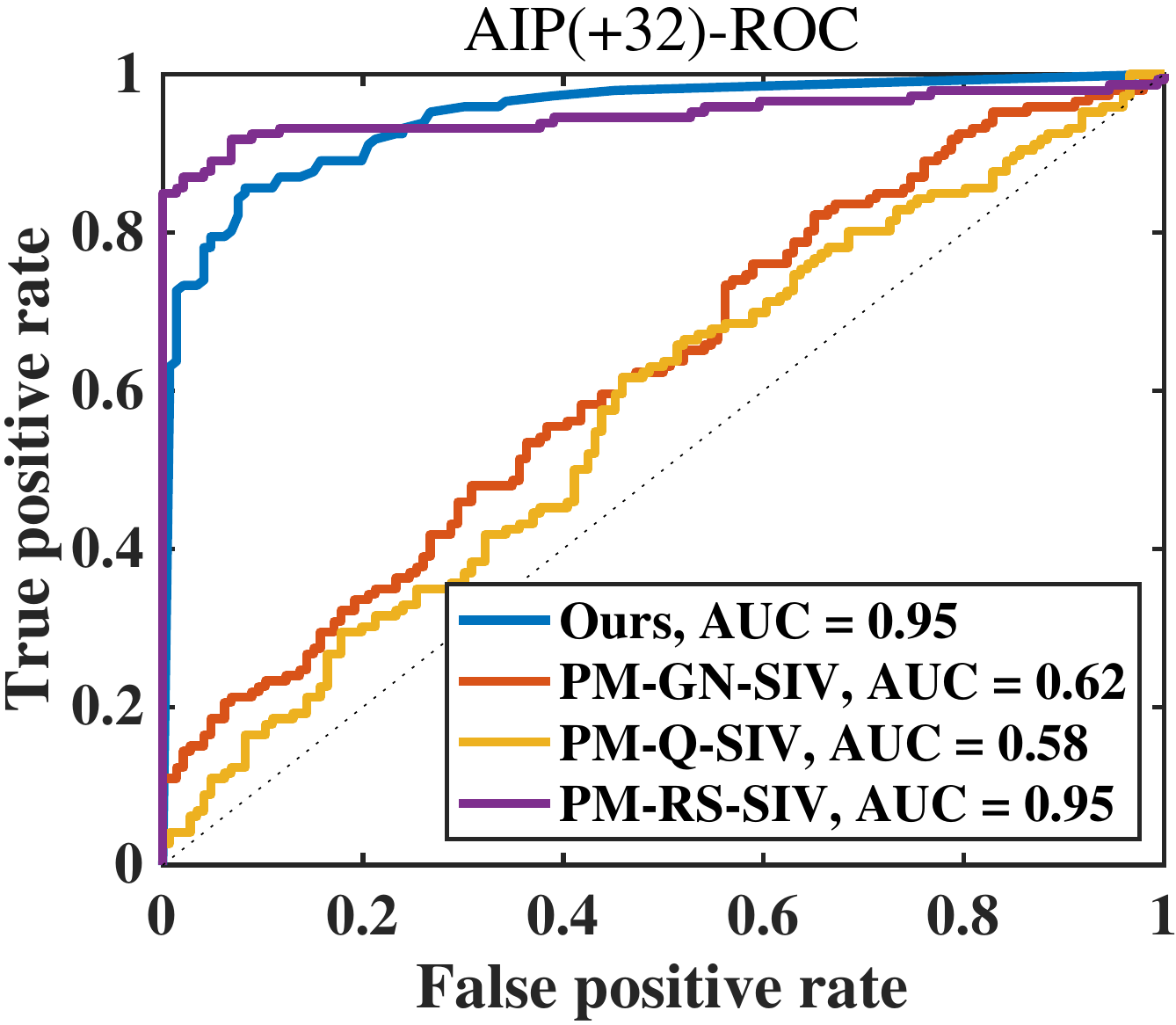}
    \end{minipage}
    \begin{minipage}{0.2\textwidth}
    \includegraphics[width=\textwidth]{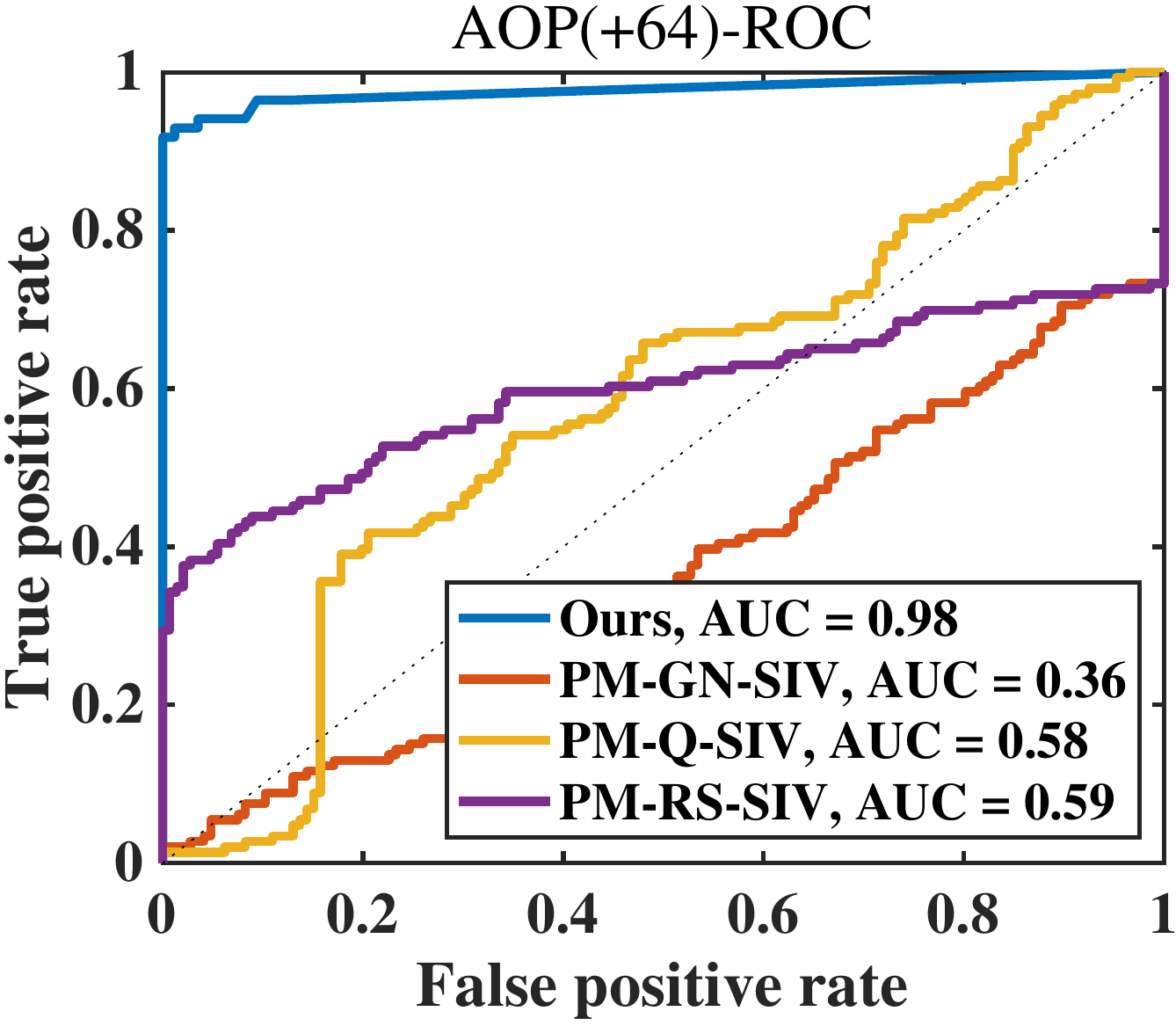}
    \end{minipage}\hspace{+1em}
    \begin{minipage}{0.2\textwidth}
    \includegraphics[width=\textwidth]{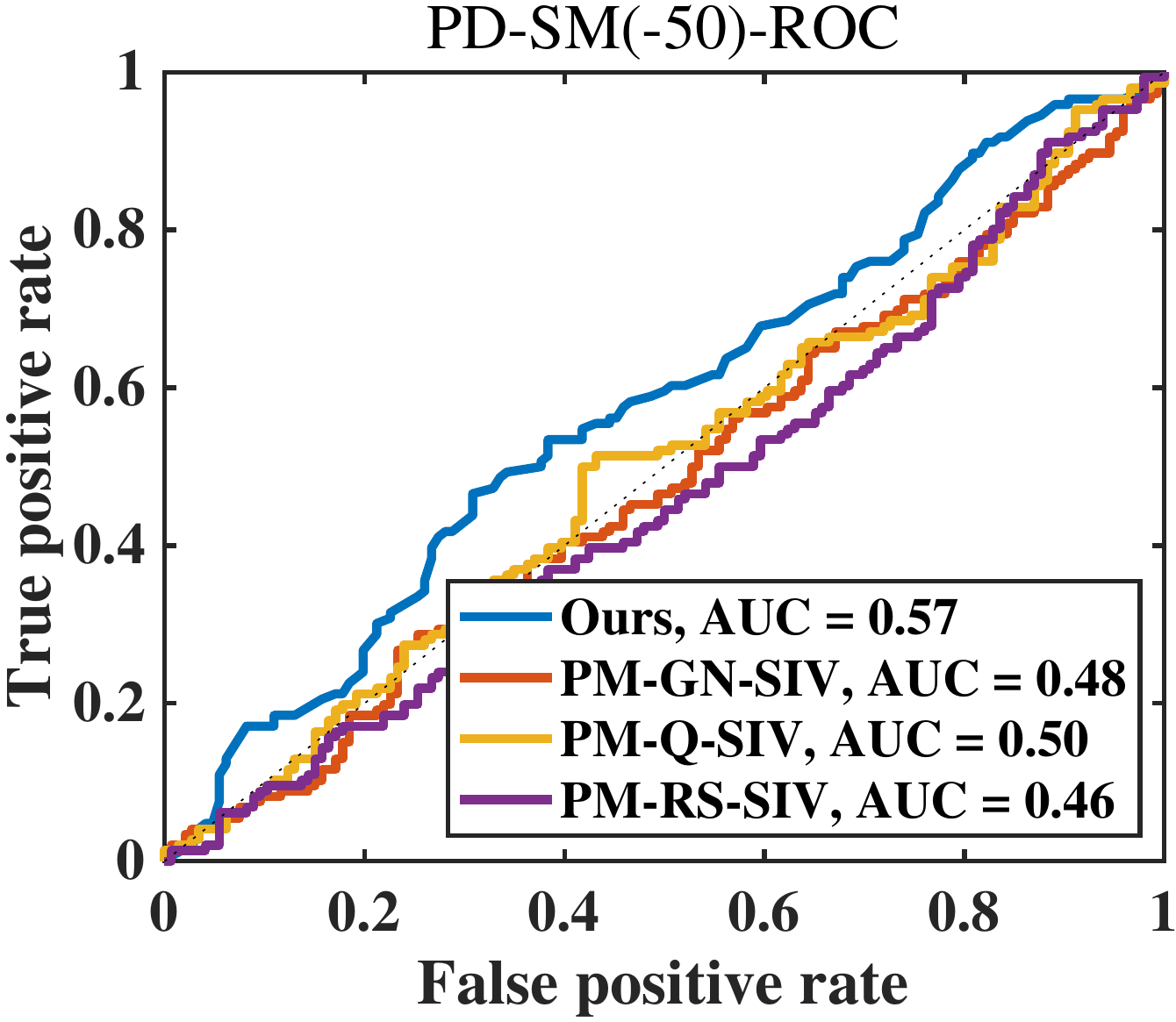}
    \end{minipage}
	\caption{Receiver operating characteristic (ROC) curves and area under ROC (AUC) of the PM series and our detector 3D-VFD on detecting different types of 3D adversarial attacks.}
    \label{fig4}
    \vspace{-1em}
\end{figure}
\subsection{Comparison with Prior Work}
\cite{yang2021adversarial} proposed a perturbation-measurement (PM) principle for 3D adversarial point cloud detection. In this experiment, we select PM equipped with three perturbation methods (namely Gaussian Noising (GN), Quantization (Q), and Random Sampling (RS)) and the best-performing measurement Set-Indiv Variance (SIV), for comparison. Simultaneously, 3D adversarial point clouds generated by Imperceptible Transfer Attack (ITA) \cite{ITA}, Shape Invariant Attack (SIA) \cite{SIA}, APP \cite{Xiang_2019_CVPR}, AIP(+32) \cite{Xiang_2019_CVPR}, Adversarial Object Perturbation with 64 points added (AOP(+64)) \cite{Xiang_2019_CVPR}, and PD-SM(-50) \cite{Saliency} are chosen as our detection objects. Note that PM has four determinant parameters and adopts different parameters for different types of adversarial attacks. Here we follow its parameter selection and detection performance evaluation metric AUC for the comparison experiments. Relevant experimental results are reported in Fig. \ref{fig4}. As we see, compared with the PM series, 3D-VFD exhibits a remarkable detection performance superiority on various types of 3D adversarial point clouds without the help of the victim model. Further, we find that 3D-VFD is better at detecting adversarial point clouds generated by point perturbation, which may be because 3D-VFD pays great attention to the local geometric information of point clouds (see Section 2.1.2). Moreover, 3D-VFD also achieves eye-catching detection results on point-adding-based attacks because adding points can be seen as a large perturbation to point clouds. However, 3D-VFD performs poorly in detecting point clouds with a small number of points removed but still has a significant advantage over the PM series.
\vspace{-0.5em}
\subsection{Time Complexity of Detection}
As described in Section 3, 3D-VFD consists of feature mapping and classification modules. To evaluate the time complexity of 3D-VFD, we choose 1460 adversarial point clouds, each containing 1024 points, for testing. Meanwhile, we also test its running time on point clouds with 2048 points. Relevant tests are carried out on a computer with Xeon E5-2630. The result is that the RGF mapping module takes 0.7901 seconds per point cloud and 2.3918 seconds per point cloud on the two tests, respectively, and the second module takes trim time. In practice, we can use parallel computing to speed up the detection rate of 3D-VFD.
\vspace{-0.5em}
\section{Conclusion}
In this paper, we rethink the defense against 3D adversarial point clouds through the lens of steganalysis. Specifically, we design a 3D adversarial point cloud detector that can effectively detect multiple types of 3D adversarial attacks while keeping fast detection speed. However, as we see, our detector relies on hand-crafted geometric features, which imposes certain limitations on our detector, especially when it is faced with point removing based attacks. In future work, we will attempt to design an end-to-end deep network for 3D adversarial point cloud detection.
\small
\bibliographystyle{IEEEbib}
\bibliography{icme2023template}
\end{document}